\documentclass[11pt]{article}
\usepackage[letterpaper,textwidth=6.5in,textheight=9in,
            centering,ignorehead,nomarginpar]{geometry}

\usepackage[T1]{fontenc}
\usepackage{textcomp}
\usepackage{xcolor}

\usepackage{setspace}

\usepackage{url}
\usepackage{hyperref}
\usepackage{doi}
\usepackage{multirow}
\usepackage{array}
\usepackage{natbib}

\bibpunct{(}{)}{;}{a}{,}{;}

\usepackage{graphicx}
\usepackage{bbm}
\usepackage{amsmath,amsthm,amssymb,amsfonts,latexsym}
\usepackage{mathtools}
\usepackage{booktabs}
\usepackage{icomma}
\usepackage{afterpage}
\usepackage{titling}
\thanksmarkseries{alph}
\usepackage[labelfont={small,sc},textfont={small,it},
            margin=15pt,skip=10pt,position=bottom]{caption}
\usepackage[labelfont={scriptsize,normalfont},textfont={scriptsize,it},
            margin=20pt,skip=5pt,position=bottom,
            labelformat=simple]{subcaption}

\usepackage[ruled,vlined,algosection]{algorithm2e}

\newtheorem{remark}{Remark}

\numberwithin{condition}{section}
\numberwithin{assumption}{section}
\numberwithin{remark}{section}
\numberwithin{equation}{section}
\numberwithin{lemma}{section}
\numberwithin{definition}{section}
\numberwithin{theorem}{section}
\numberwithin{proposition}{section}
\numberwithin{table}{section}
\numberwithin{figure}{section}
\numberwithin{theorem}{section}
\numberwithin{corollary}{section}
\numberwithin{property}{section}

\def\n1{n}

\newsavebox{\savepar}

\numberwithin{equation}{section}
\numberwithin{table}{section}
\numberwithin{figure}{section}

\newcommand{\myred}[1]{{\color{black}{#1}}}

\usepackage[running]{lineno}

\begin{document}
\title{
Making Leveraged Exchange-Traded Funds Work for your Portfolio 
}
\author{Peter A. Forsyth\thanks{Cheriton School of Computer Science, University of Waterloo, Waterloo
ON, Canada, N2L 3G1, \texttt{paforsyt@uwaterloo.ca}} \and Pieter M. van Staden\thanks{National Australia Bank, Melbourne, Victoria, Australia 3000. The
research results and opinions expressed in this paper are solely those
of the authors, are not investment recommendations, and do not reflect
the views or policies of the NAB Group. \texttt{pieter.vanstaden@gmail.com}} \and Yuying Li\thanks{Cheriton School of Computer Science, University of Waterloo, Waterloo
ON, Canada, N2L 3G1, \texttt{yuying@uwaterloo.ca}} }

\date{June 23, 2025}

\maketitle


\begin{abstract}
We examine strategically incorporating broad stock market leveraged exchange-traded funds
(LETFs) into investment portfolios.
We demonstrate
that easily understandable and implementable
strategies can enhance the risk-return profile of a portfolio containing LETFs. 
Our analysis shows that seemingly reasonable investment strategies
may result in undesirable Omega ratios, with these effects compounding
across rebalancing periods. By contrast, relatively simple dynamic
strategies that systematically de-risk the portfolio once gains are
observed can exploit this compounding effect, taking advantage of
favorable Omega ratio dynamics. Our findings suggest that LETFs represent
a valuable tool for investors employing dynamic strategies, while
confirming their well-documented unsuitability for passive or static
approaches.

\vspace{5pt}


\end{abstract}

\section{Introduction}

Leveraged Exchange Traded Funds (LETFs) are exchange-traded funds
(ETFs) replicating some multiple $\beta$ of the daily returns of
their underlying reference assets or indices before costs. In contrast
with 'vanilla' ETFs (VETFs), which simply aim to replicate the returns
of their underlying assets/indices before costs (i.e., $\beta=1$),
typical LETF multipliers are $\beta=2$ or $\beta=3$ of daily returns
in the case of leveraged long exposure\footnote{Inverse LETFs, with negative return multipliers, are not considered
in this paper}. 

Despite considerable skepticism within academic circles, much of it
warranted given the inherent risks, the persistent and growing popularity
of LETFs among both retail and institutional investors suggests these
market participants appreciate opportunities that academic analysis
may overlook, pointing to a more nuanced risk-reward dynamic (see
\citet{van_staden_2025} for a detailed discussion).

\begin{figure}[htb!]
\centerline{\begin{subfigure}[t]{.48\linewidth} \centering \includegraphics[width=1\linewidth]{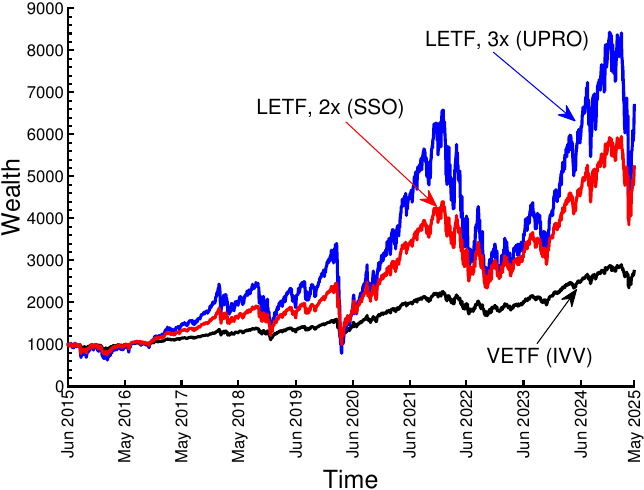}
\caption{S\&P 500: LETFs and VETF}
\label{fig_Actual_ETFs_SP500} \end{subfigure}\qquad{}\begin{subfigure}[t]{.48\linewidth}
\centering \includegraphics[width=1\linewidth]{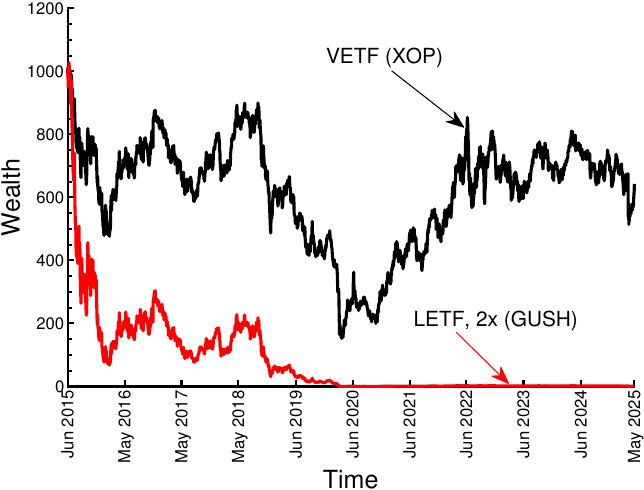}
\caption{S\&P Oil \& Gas Exploration \& Production Select Industry Index: VETF
and LETF }
\label{fig_Actual_ETFs_OilGas} \end{subfigure} } \caption{LETFs written on broad stock market indices vs. LETFs on niche sector
indices: Illustrative buy-and-hold portfolio values from investing
all hypothetical initial wealth of \$1000 in each VETF and LETF from
June 2015 until May 2025. Please note that we are \textit{not} making
the case for simple buy-and-hold strategies in this paper - these
figures are intended to illustrate the behavior of the LETF relative
to its corresponding VETF over time. Returns have been adjusted for
splits and dividends, data from nasdaq.com. }
\label{fig_Actual_ETFs} 
\end{figure}

Crucially, our focus is limited to LETFs written on broad stock market
indices---a critical qualification. LETFs based on broad, diversified
indices like the S\&P 500 reference an ``underlying asset'' (i.e.
the market index itself) which has high levels of diversification,
relatively lower volatility, and a tendency to exhibit positive long-term
drift. For example, Figure \ref{fig_Actual_ETFs_SP500} considers
a simple buy-and-hold position for three ETFs referencing the S\&P
500 as underlying asset: A standard VETF replicating the index (IVV),
a LETF with daily returns multiplier $\beta=2$ (SSO), and a LETF
with daily returns multiplier $\beta=3$ (UPRO). In contrast, Figure
\ref{fig_Actual_ETFs_OilGas} illustrates that in the case of the
S\&P Oil \& Gas Exploration \& Production Select Industry Index, which
has about 50 constituents in a volatile sector, simple buy-and-hold
positions in the VETF (XOP) and the corresponding LETF with multiplier
$\beta=2$ (GUSH) show how LETFs can indeed live up to their bad reputation. 

However, we emphasize that we are \textit{not} making the case for
simple buy-and-hold strategies involving LETFs. Instead, we present
relatively sophisticated yet easily understandable and implementable
strategies that require only infrequent rebalancing. Figure \ref{fig_Actual_ETFs}
is only included as an illustration for why our focus remains on LETFs
referencing broad stock market indices, rather than LETFs on niche
sector indices. 

\noindent 

So how does one go about incorporating broad stock market LETFs into
a portfolio, to take advantage of relative return behavior observed
in Figure \ref{fig_Actual_ETFs_SP500}? The contributions of this
paper are as follows:
\begin{itemize}
\item We avoid the trap of using only available LETF market data, since
LETFs were only introduced in 2006 (\citet{BansalMarshall2015}) and
therefore cannot provide a thorough historical perspective of LETF
behaviour during different market and inflation regimes\footnote{Many LETFs were only introduced much later. For example, 
the LETF
GUSH illustrated in Figure \ref{fig_Actual_ETFs_OilGas} was only
introduced in May 2015.}. Figure \ref{fig_Actual_ETFs_SP500} might reflect only the relatively
benign market conditions (with notable exceptions) of the last decade,
and we do not want to assume that broad stock market LETFs will always
have the behavior relative to the VETF as observed in Figure \ref{fig_Actual_ETFs_SP500}.
Instead, we construct a proxy returns time series for a VETF and LETF
referencing a broad equity market index with data since 1926, to obtain
a truly long-term, robust perspective on introducing LETFs to a portfolio.
The synthetic time series appropriately incorporates ETF costs and
interest, and is inflation-adjusted to enable realistic conclusions.
\item Intuitive explanations are provided as to why LETFs can enhance portfolio
performance when used appropriately within the context of a dynamic
strategy, which offers accessible insights for practitioners. 
\item The critical importance of the Omega ratio in evaluating medium-term
LETF strategies is demonstrated, revealing how this metric captures
performance dynamics that traditional risk measures may miss. Our
analysis shows that while some seemingly reasonable LETF strategies
exhibit Omega ratios below unity, others achieve ratios above unity,
with compounding effects across multiple rebalancing periods. We show
how systematic de-risking following gains allows investors to capitalize
on favorable Omega ratio compounding. This provides the intuitive
explanation for the superior path-dependent performance observed in
the literature (\citet{van_staden_2025}). 
\item Finally, we construct optimal dynamic LETF investment strategies using
a data driven neural network approach, which relies only on historical
returns data, i.e. no parametric models are assumed for the return
dynamics of underlying assets. The results confirm the benefits and
risks, as well as significant improvements to Omega ratios, that can
be realized by incorporating LETFs into portfolios. 
\end{itemize}
While our findings strongly support the use of LETFs within dynamic,
actively managed strategies, they simultaneously reinforce warnings
against their use in passive or static investment contexts. 

The paper is structured as follows. Section \ref{jump_section} provides
intuition under the assumption of parametric asset dynamics where
the underlying stock index follows a jump diffusion process calibrated
to historical stock market data, with corresponding illustrative investment
results in Section \ref{sec:Jump-Diffusion results}. Building on
these insights, Section \ref{sec:Fixed weight simulations} demonstrates
the investment outcomes obtained by rebalancing to different fixed
LETF weight allocations. Section \ref{sec:Data-driven-Machine-Learning}
generalizes these results by using a data-driven neural network approach
to determine the optimal dynamic allocation of wealth to the LETF
at each rebalancing time, confirming the contrarian nature of strategies
designed to take advantage of LETF behavior. Section \ref{sec:Conclusion}
concludes the paper.

\section{Intuition: jump diffusion} \label{jump_section}
While the ultimate goal is to show LETF investment strategies without
any parametric assumptions regarding the dynamics of the underlying
assets (see Section \ref{sec:Data-driven-Machine-Learning}), to gain
the necessary intuition it is nevertheless instructive to start with
simple parametric dynamics for the underlying assets.

We first consider the basic case where the underlying stock index
follows a jump diffusion.  For the readers convenience, we highlight key elements of
the derivation in \cite{AvellanedaZhang2010,Ahn_2015,van_staden_2025}.

Let the value of the  stock index be denoted by $S(t)$ and the value
of a risk free bond be denoted by $B(t)$.
Assume that the stock index follows a jump diffusion process, which
allows for non-normal returns.
If a jump occurs  $S(t) = \xi S(t^-)$, and
\begin{eqnarray}
  \frac{dS}{S(t^-)} & = & ( \mu - \lambda \kappa) ~dt + \sigma~dZ + (\xi-1) d\mathbb{Q} \nonumber \\
                 & & d\mathbb{Q} = \begin{cases}
                          0 & ;~ {\text{ probability }} (1 - \lambda ~dt) \\
                          1 & ;~ {\text{ probability }} \lambda ~dt 
                     \end{cases} \nonumber \\
            & & \kappa = E[\xi-1] \nonumber \\
            & & \lambda = {\text{ intensity of the Poisson process}} \nonumber \\
            & & dZ = {\text{ increment of a Wiener process}}
      ~.
     \label{S_sde_jumps}
\end{eqnarray}
Assume that $y= \log \xi$ follows a double exponential process\citep{kou:2002},
with density $g(y)$ given by
\begin{eqnarray}
    g( y ) & = &   p_{up} \eta_1 e^{-\eta_1 y} {\bf{1}}_{y \geq 0} +
       (1-p_{up}) \eta_2 e^{\eta_2 y} {\bf{1}}_{y < 0}.
\label{eq:dist}
\end{eqnarray}
where $p_{up}$ is the probability of an upward jump.
Note as well that
\begin{eqnarray}
E[\xi] &= \frac{p_{\text{\emph{{up}}}}\eta_1}{\eta_1 - 1} +
          \frac{(1-p_{\text{\emph{{up}}})}\eta_2}{\eta_2 + 1} ~.
\end{eqnarray}
Equation (\ref{S_sde_jumps}) implies that
\begin{eqnarray}
    \frac{S(t)}{S(0)} & = & 
            e^{(\mu - \lambda \kappa) t + \sigma (Z(t) - Z(0))} e^{- \sigma^2 t/2 + \sum_{i=0}^{\mathbb{N}(t)} \log \xi_i}
   ~,
             \label{S_exact_jump}
\end{eqnarray}
where $\mathbb{N}(t)$ counts the number of Poisson jumps with intensity $\lambda$ in $(0,t)$,
and $\xi_i$ are drawn from the density (\ref{eq:dist}), and 
$(Z(t) - Z(0) ) \simeq \mathcal{N} (0, t)$
where $\mathcal{N} (0, t)$ is a draw from a normal distribution with mean zero and variance $t$.

The bond  is considered to be risk-free and non-volatile
\begin{eqnarray}
  dB & = & r B~dt ~. \label{bond_eqn}
\end{eqnarray}

{\myred{
In the absence of limited liability, 
the value of a leveraged ETF  $V^{\ell}(t)$, which is continuously rebalanced to a weight of $\beta$
in the stock index and $(1-\beta)$ in the bond index then follows the process
(we assume $\beta >1$)
\begin{eqnarray}
\frac{dV^{\ell}}{V^{\ell}} & = &  \beta \biggl( \frac{dS}{S} \biggr) 
               + (1 - \beta) \biggl( \frac{dB}{B} \biggr)  - c_{\ell}~dt \nonumber \\
              & = &\biggl( (1- \beta) r + \beta (\mu - \lambda \kappa) -c_\ell \biggr) ~dt 
                      + \beta \sigma ~dZ + \beta( \xi -1) ~d\mathbb{Q} \label{letf_sde_1}
     \label{letf_sde_2}        ~,
\end{eqnarray}
where $c_{\ell}$ is the leveraged ETF (denoted by LETF) expense ratio.  
Since we assume continuous rebalancing of the LETF, $V^{\ell}$ can only become negative due
to jumps. 
Hence, to incorporate limited liability,  i.e. the value of the LETF cannot become negative \citep{Ahn_2015,van_staden_2025},
we can rewrite equations \eqref{letf_sde_2} to reflect this condition
\begin{eqnarray}
     \frac{dV^{\ell}}{V^{\ell}} & = & 
             \biggl( (1- \beta) r + \beta (\mu - \lambda \kappa) -c_\ell \biggr) ~dt 
                      + \beta \sigma ~dZ + \max( \beta( \xi -1), -1) ~d\mathbb{Q} ~,
       \label{letf_sde_3}
\end{eqnarray}
which implies that
\begin{eqnarray}
     \frac{V^{\ell}(t)}{V^{\ell}(0)} & = &
             e^{  -c_\ell t }  
                  e^{ \left( ( 1 - \beta) r  + \beta (\mu - \lambda \kappa)  - \beta^2 \sigma^2/2 \right) t  } 
                  e^{ \beta \sigma (Z(t) - Z(0)) 
                     + \sum_{i=0}^{\mathbb{N}(t)}  \log \left( \max( 1 + \beta( \xi_i-1), 0) \right) } \nonumber \\
             & =& e^{ -c_\ell t }  e^{ \left( 1 - \beta) r  - \beta^2 \sigma^22  \right)t } 
                 \biggl( e^{(\mu - \lambda \kappa) t + \sigma (Z(t) - Z(0)) } \biggr)^\beta
                  e^{ \sum_{i=0}^{\mathbb{N}(t)}  \log \left( \max( 1 + \beta( \xi_i-1), 0) \right) }
    ~. \nonumber \\
             \label{letf_exact_jump}
\end{eqnarray}
We note that, in practice, LETFs are often replicated using index swaps \citep{Guasoni_2023}.  This is, of course,
economically equivalent to equation (\ref{letf_sde_1}). 
}
}

Rewrite equation (\ref{S_exact_jump}) as
\begin{eqnarray}
  \biggl( \frac{S(t)}{S(0)} \biggr)^{\beta} e^{ \beta \sigma^2 t/2 -\beta \sum_{i=0}^{\mathbb{N}(t)} \log \xi_i}
    & = & \biggl( e^{(\mu - \lambda \kappa) t + \sigma (Z(t) - Z(0))} \biggr)^{\beta}
   ~.
     \label{S_exact_jump_2}
\end{eqnarray}
Substitute equation (\ref{S_exact_jump_2}) into equation (\ref{letf_exact_jump}) to obtain
\begin{eqnarray}
    \frac{V^{\ell}(t)}{V^{\ell}(0)}  & = &  e^{\left\{ 
                          \left(  (1-\beta)r + \beta (1-\beta) \sigma^2 /2 -c_{\ell} \right) t 
                              \right\} }
                   \biggl( \frac{S(t)}{S(0)} \biggl)^{\beta} H(\beta,t) ~,
\end{eqnarray}
where
\begin{eqnarray}
     H(\beta,t) & = & \Pi_{i=0}^{\mathbb{N}(t)}  \biggl(  \frac{ \max( 1 + \beta(\xi_i-1), 0) }{\xi_i^\beta }
                                                        \biggr)  ~.
\end{eqnarray}

Now consider a portfolio containing an initial allocation of $\alpha^{\ell}$ to the LETF and
$(1-\alpha^\ell)$ to the risk free bond, with $0 \leq \alpha^\ell \leq 1$.  Given an initial
wealth $W(0)$, the total portfolio 
value at time $t$, denoted by $P^\ell(t)$, is then
\begin{eqnarray}
    \frac{P^{\ell}(t)} { W(0)} & = & (1 - \alpha^\ell) \left( \frac{B(t)}{B(0)} \right)
                                         + \alpha^\ell \left( \frac{V^{\ell}(t)}{V^{\ell}(0)} \right) \nonumber \\
                        & = & (1 - \alpha^l) e^{rt} +
                       \alpha^l  e^{ \left\{ \left( (1-\beta)r 
                           + \beta (1-\beta) \sigma^2 /2 -c_{\ell}  \right) t  \right\} }
                                   \biggl( \frac{S(t)}{S(0)} \biggl)^{\beta}  H(\beta,t) ~.
    \label{letf_portfolio}
\end{eqnarray}
The value of a Vanilla ETF $V^{{v}}$ is simply (with the same initial wealth $W(0)$)
\begin{eqnarray}
\frac{V^{{v}}(t)}{W(0)}  & = &  e^{-c_v t} \left( \frac{S(t)}{S(0)} \right) ~,
\end{eqnarray}
where $c_v$ is the expense ratio of the vanilla ETF (denoted by VETF).  Consider a portfolio containing an initial allocation of 
$\alpha^{{v}}$ to the VETF and
$(1-\alpha^{v})$ to the risk free bond, with $0 \leq \alpha^{v} \leq 1$.  The total portfolio
value at time $t$ is then
\begin{eqnarray}
    \frac{P^{{v}}(t)} { W(0)} & = & 
                         (1 - \alpha^{v}) e^{rt} + \alpha^{v} e^{-c_v t} \biggl( \frac{S(t)}{S(0)} \biggl)~,
    \label{vetf_portfolio}
\end{eqnarray}
assuming initial wealth $W(0)$.

\begin{remark}[Properties of equation (\ref{letf_portfolio})] \label{drag_remark}
It is useful to note the following properties of  equation (\ref{letf_portfolio})
\begin{itemize}
  \item $e^{ \left\{ \left( (1-\beta)r 
                           + \beta (1-\beta) \sigma^2 /2 -c_{\ell}  \right) t  \right\} } < 1$ 
            ; if $\beta > 1, t>0$,
   \item $H(\beta,t) <1$ ; if $\beta > 1, t>0$,\citep{van_staden_2025},
 \end{itemize}
which implies that volatility, jumps and expenses act as a drag on the LETF.
However, the power law term $(S(t)/S(0))^\beta$ counteracts this drag if $(S(t)/S(0)) > 1$ ($\beta >1$).
\end{remark}

\subsection{GBM case}
In order to gain some intuition, we first consider a case of geometric Brownian motion 
(GBM) dynamics. This can be formally obtained
from the results in Section \ref{jump_section} by setting the jump intensity $\lambda = 0$.

More precisely, we obtain
\begin{eqnarray}
   \frac{P^{\ell}(t)} { W(0)} & = & 
                        (1 - \alpha^l) e^{rt} +
                       \alpha^l  e^{ \left\{ \left( (1-\beta)r 
                           + \beta (1-\beta) \sigma^2 /2 -c_{\ell}  \right) t  \right\} }
                                   \biggl( \frac{S(t)}{S(0)} \biggl)^{\beta}  ~, \label{letf_gbm}
                 \\
    \frac{P^{{v}}(t)} { W(0)} & = & 
                         (1 - \alpha^{v}) e^{rt} + \alpha^{v} e^{-c_v t} \biggl( \frac{S(t)}{S(0)} \biggl)
              ~,
    \label{vetf_gbm} \\
   \frac{S(t)}{S(0)} & = & 
             e^{(\mu  - \sigma^2/2)t} e^{\sigma (Z(t) - Z(0)) }
   ~.
             \label{S_gbm}
\end{eqnarray}

We use data from the Center for Research in Security Prices (CRSP)
on a monthly basis over the 1926:1-2023:12 period.\footnote{More
specifically, results presented here were calculated based on data from
Historical Indexes, \copyright 2023 Center for Research in Security
Prices (CRSP), The University of Chicago Booth School of Business.
Wharton Research Data Services (WRDS) was used in preparing this
article. This service and the data available thereon constitute valuable
intellectual property and trade secrets of WRDS and/or its third-party
suppliers.}
Our base case tests use the CRSP US 30 day T-bill for the
bond asset and the CRSP value-weighted total return index for the stock
asset. This latter index includes all distributions for all domestic
stocks trading on major U.S.\ exchanges. All of these various indexes
are in nominal terms, so we adjust them for inflation by using the U.S.\
CPI index, also supplied by CRSP. We use real indexes since investors
should be focused on real (not nominal)
wealth goals.  The parameters in Table \ref{data_table} are obtained
using maximum likelihood.

\begin{table}[hbt!]
\begin{center}
\begin{tabular}{lc} \hline
$\mu$ &  0.0818\\
 $\sigma$ &  .1849 \\
 LETF leverage & $\beta = 2$\\
30 day T-bill return $r$ & 0.0032\\
LETF expense ratio $c_\ell$ & .0089 \\
VETF expense ratio $c_v$ & 0.0 \\
\hline
\end{tabular}
\caption{Data for GBM example.  Annualized parameters fit to 
CRSP monthly data, 1926:1-2023:12, inflation adjusted.
The LETF expense ratio corresponds to that of the ProShares Ultra S\&P 500 LETF with multiplier $\beta = 2$ (etfdb.com/etf/SSO, accessed 15 May 2025).
}
\label{data_table}
\end{center}
\end{table}

We will assume that the leverage factor for the LETF is $\beta = 2$.  After an initial
allocation to the risk free bond and the LETF/VETF, there is no further rebalancing,
with an initial investment horizon of $T=1.0$ years.

\begin{figure}[htb!]
\centerline{%
\begin{subfigure}[t]{.40\linewidth}
\centering
\includegraphics[width=\linewidth]{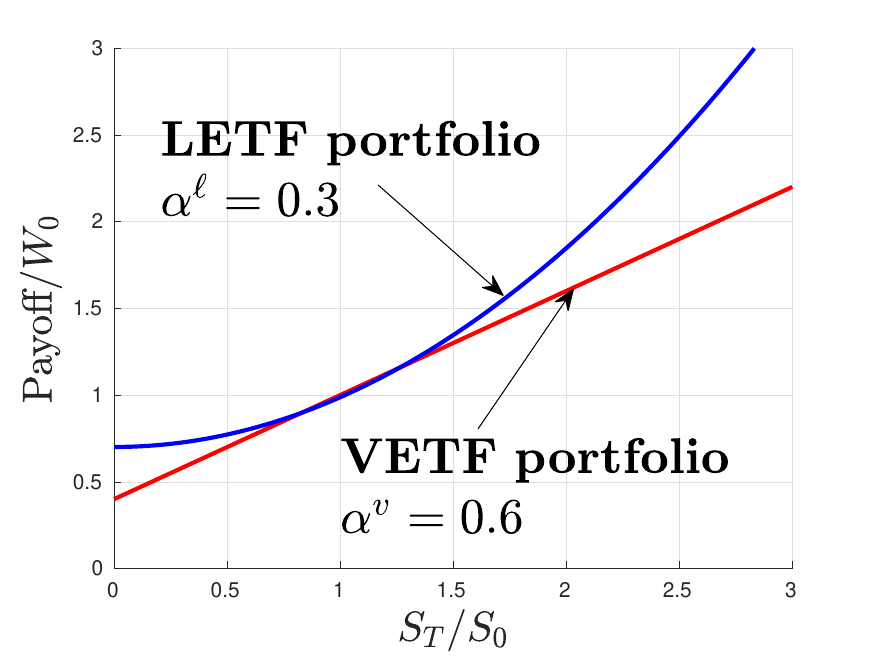}
\caption{$P^{\ell}/W(0), P^{v}/W(0)$}
\label{simple_payoff_three}
\end{subfigure}
\begin{subfigure}[t]{.40\linewidth}
\centering
\includegraphics[width=\linewidth]{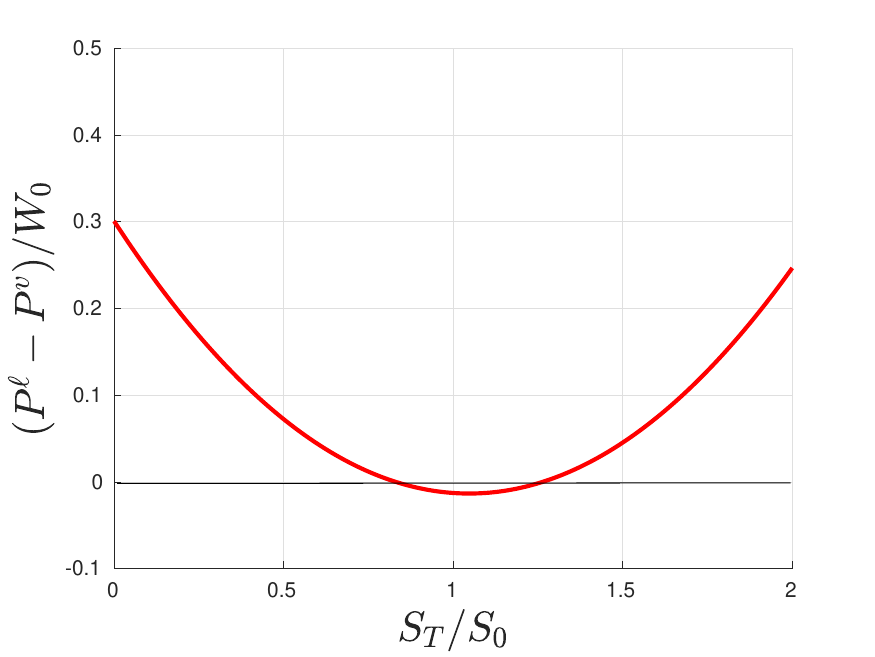}
\caption{$(P^{\ell} - P^{v})/W(0)$ from Figure \ref{simple_payoff_three}}
\label{payoff_zoom_three}
\end{subfigure}
}
\caption{
Payoff diagrams comparing $P^{v}/W(0), P^{\ell}/W(0)$. $T=1.0$ years,
from equations (\ref{letf_gbm}-\ref{S_gbm}), GBM case.
LETF leverage factor $\beta = 2.0$.
Stock data fit of equation (\ref{S_gbm}) to inflation adjusted CRSP index, 1926:1-2023:12.
Data in Table \ref{data_table}.
$\alpha^{\ell} = 0.3, \alpha^v = 0.6$.
}
\label{payoff_fig_point_three}
\end{figure}

Figure \ref{payoff_fig_point_three} shows the payoff diagrams from equations (\ref{letf_gbm}-\ref{S_gbm}),
assuming that the initial allocation fraction to the LETF for $P^{\ell}$ is $\alpha^\ell = 0.30$,
compared the $\alpha^v = 0.60$, for $P^{v}$ (the portfolio which uses a vanilla stock ETF).
Since the leverage ratio for the LETF is $\beta = 2.0$,  this allocation to the LETF
in $P^\ell$ results in the same initial exposure to stock price changes  as for the Vanilla
ETF portfolio $P^v$.  

Figure \ref{payoff_fig_point_three} clearly shows the nonlinear payoff effect of using an LETF compared
with using a VETF.  Figure \ref{payoff_zoom_three} focuses on the difference $(P^{\ell} - P^v$).  We can
see that use of an LETF is a drag on performance near $S_T/S_0 \simeq 1$, but boosts performance
for either large or small stock returns evaluated over the entire year.
{\myred{The enhanced payoff of the portfolio containing the LETF if $(S_T/S_0) <1$ is
due to the larger allocation to bonds, compared to the portfolio with the VETF.}}

\begin{figure}[htb!]
\centerline{%
\begin{subfigure}[t]{.40\linewidth}
\centering
\includegraphics[width=\linewidth]{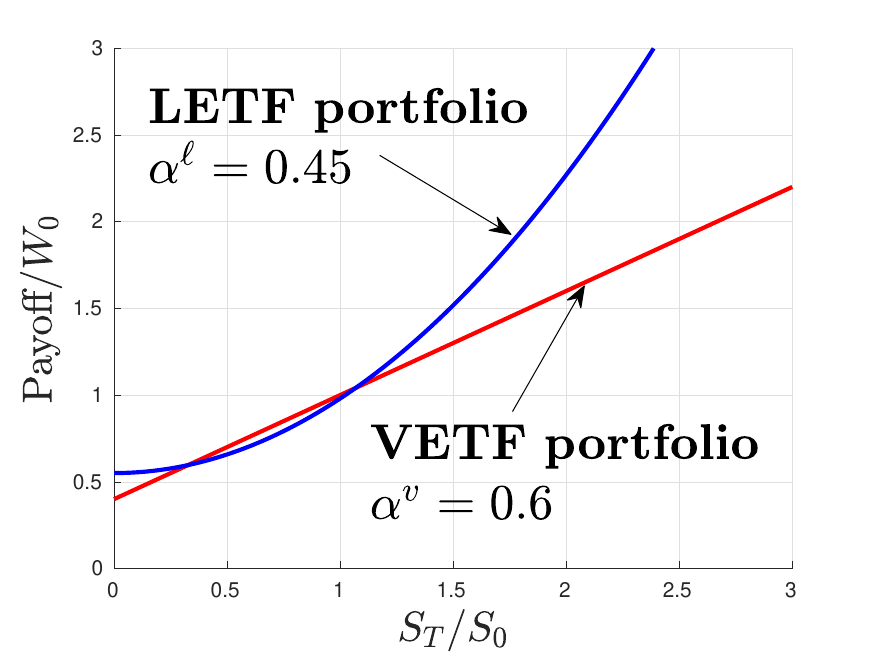}
\caption{$P^{\ell}/W(0), P^{v}/W(0)$}
\label{simple_payoff_four_five}
\end{subfigure}
\begin{subfigure}[t]{.40\linewidth}
\centering
\includegraphics[width=\linewidth]{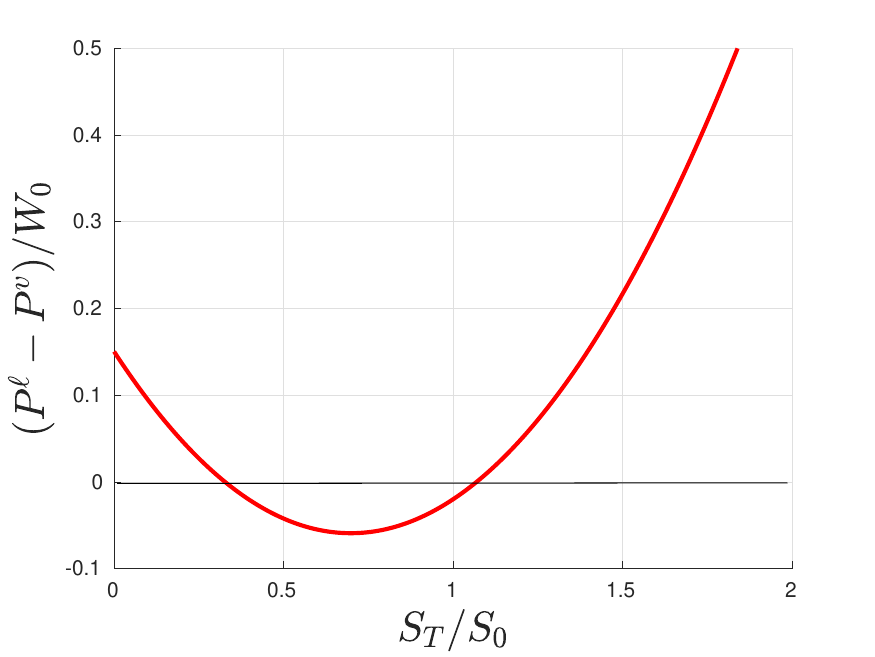}
\caption{$(P^{\ell} - P^{v})/W(0)$ from Figure \ref{simple_payoff_four_five}}
\label{payoff_zoom_four_five}
\end{subfigure}
}
\caption{
Payoff diagrams comparing $P^{v}/W(0), P^{\ell}/W(0)$. $T=1.0$ years,
from equations (\ref{letf_gbm}-\ref{S_gbm}), GBM case.
LETF leverage factor $\beta = 2.0$.
Stock data fit of equation (\ref{S_gbm}) to inflation adjusted CRSP index, 1926:1-2023:12.
Interest rate $r$ from T-bills, inflation adjusted, 1926:1-2023:12.
Data in Table \ref{data_table}.
$\alpha^{\ell} = 0.45, \alpha^v = 0.6$.
}
\label{payoff_fig_point_four_five}
\end{figure}

Figure \ref{payoff_fig_point_four_five} shows comparable results for $\alpha^\ell = 0.45$ with
$\alpha^v = 0.60$.  From Figure \ref{payoff_zoom_four_five} we can see the underperformance of
$P^{\ell}$ (the LETF portfolio) relative to the VETF portfolio $P^v$ is now larger (compared
to $\alpha^\ell = 0.30$), but has shifted to the zone $S_T/S_0 < 1$.  Note as well that
for $S_T/S_0 > 1$, the VETF portfolio has a very rapid increase in outperformance.

We can see from Figures \ref{payoff_zoom_three} and \ref{payoff_zoom_four_five},
that the area of underperformance for $\alpha^\ell = 0.45$ is larger compared
to $\alpha^\ell = 0.30$.  
However, at this point, we cannot say anything more about the performance of these choices for
$\alpha^\ell$ unless we know the probability distribution of $S_T/S_0$.

In order to generate various
statistics for $R(T) =  P^{\ell}(T)/P^v(T)$
we will use Monte Carlo simulation.
Observe that this is a pathwise test for outperformance compared to the benchmark
VETF portfolio.

Table \ref{MC_stats} shows the summary statistics for these simulations.
ES(5\%) is the expected shortfall at the five per cent level, i.e.
the mean of the worst 5\% of the outcomes.
The Omega ratio \citep{Keating_2002} at level $L$ is defined as
\begin{eqnarray}
   \Omega(L) & = & 
                        \frac{\int_L^{\infty} \left( 1- F(R_T) \right)~dR_T}
                             {\int_{-\infty}^L F(R(T)~dR_T}    ~;~ F(R_T) {\text{ CDF of }}  R_T \nonumber \\
                      & = &   \frac{E[ \max( R_T - L, 0)]}{ E[\max( L -R_T, 0) ]}
            \nonumber \\
                  & & ~~~R_T = \frac{ P^{\ell}_T}{P^{v}_T} ~.  \label{omega_def}
\end{eqnarray}
The Omega ratio is a measure of upside versus downside, with respect to the level L.  Since
we are interested in pathwise outperformance of the LETF portfolio compared to the VETF
portfolio, we will examine $\Omega(1)$.

Table \ref{MC_stats} shows that the LETF $\alpha^\ell = .45$ portfolio 
has $E[R_T]  > 1 $ and $Median[R_T]  \simeq 1$,
in contrast to the underperformance of the mean and median (relative to the VETF portfolio) 
for the $\alpha^\ell =.30$ portfolio.  Of course, this comes at a cost, since the 
expected shortfall $ES(5\%)$ and the $5'th$ percentiles are worse for $\alpha^\ell = .45$ compared
to $\alpha^v=.30$.

However, $\Omega(1) = 1.82$ for $\alpha^\ell = 0.45$ compared to $\alpha^\ell = 0.30$, 
which has $\Omega(1) = 0.71$.  More intuitively, $\alpha^\ell=.45$ has a significant upside
compared to the downside (in terms of outperformance).   In contrast, $\alpha^\ell = 0.30$ has
more downside compared to upside.

{\small
\begin{table}[hbt!]
\begin{center}
\begin{tabular}{lccccccc} \toprule[1pt]
   & E[$R_T$] &  Median[$R_T$] & $R_T: 5^{th}$  & $R_T: 95^{th}$   & ES( 5\% )& $\Omega(1)$ &$Prob[R_T > 1]$\\
   &          &                &percentile & percentile &      &  & \\
    \hline
$\alpha^{\ell} = 0.3$ &  0.9981 (.0001) & .9919 & .9871 & 1.0294  & .9871  & 0.7130 & .2739  \\
\hline
$\alpha^{\ell} = 0.45$ & 1.0145 (.0004) & .9995 & .9367 & 1.1426  & .9315  & 1.8247 & .4978  \\
\bottomrule[1pt]
\end{tabular}
\caption{
Statistics for $R(T) = P^{\ell}(T)/P^v(T)$. $\alpha^v = 0.6$.
$T=1.0$ years,
from equations (\ref{letf_gbm}-\ref{S_gbm}), GBM case.
LETF leverage factor $\beta = 2.0$.
Numbers in brackets are the standard error estimate at the 95\% confidence level.
Stock data fit of equation (\ref{S_gbm}) to inflation adjusted CRSP index, 1926:1-2023:12.
Interest rate $r$ from T-bills, inflation adjusted, 1926:1-2023:12.
Data in Table \ref{data_table}.   $10^5$ MC simulations.
ES(5\%) is the mean of the worst 5\% of the outcomes.
$\Omega(1)$ defined in equation (\ref{omega_def}).
}
\label{MC_stats}
\end{center}
\end{table}
}

Table \ref{MC_stats_ten} examines a more long term strategy.  We consider an investment
horizon of ten years. Initially, and annually thereafter, the VETF and LETF strategies
are rebalanced with $\alpha^v$ and $\alpha^{\ell}$ fractions in the stock ETF.
For the $\alpha^\ell=.30$ case, E[$R_T$] and Median[$R_T$] are less than one, 
with $\Omega(1) \simeq 0.40$.  For this value of $\alpha^{\ell}$, the LETF portfolio
has does not seem to be worthwhile: there is a consistent drag on performance
relative to the VETF portfolio, and the Omega ratio indicates pronounced
downside.  Note that the Omega ratio is reduced compared to the one year
case (Table \ref{MC_stats}).

In contrast, for the LETF portfolio (ten year
case, Table \ref{MC_stats_ten}) with $\alpha^{\ell} = .45$, both E[$R_T$] Median[$R_T$] are greater
than one (indicating outperformance) and with $\Omega(1) \simeq 6.4$, indicating significant
upside.  Note that Omega for the ten year strategy is much larger than for the one year
case, indicating that the Omega ratio is compounded by repeated rebalancing.
Of course, there is no free lunch here, the 5th percentile and ES(5\%) is worse
for $\alpha^{\ell} = .45$ compared to $\alpha^v = 0.30$.

{\small
\begin{table}[hbt!]
\begin{center}
\begin{tabular}{lccccccc} \toprule[1pt]
   & E[$R_T$] &  Median[$R_T$] & $R_T: 5^{th}$  & $R_T: 95^{th}$   & ES( 5\% )& $\Omega(1)$ &$Prob[R_T > 1]$\\
   &          &                &percentile & percentile &      &  & \\
    \hline
$\alpha^{\ell} = 0.3$ & .9808 (.0003 )   & .9716 & .9158  &  1.0766  & .9077   & 0.39891  & .2910  \\
\hline
$\alpha^{\ell} = 0.45$ & 1.152 (.001)   & 1.1142   & .8324    &  1.5992    & .7822   & 6.3933  & .7136   \\
\bottomrule[1pt]
\end{tabular}
\caption{
Statistics for $R(T) = P^{\ell}(T)/P^v(T)$. $\alpha^v = 0.6$.
$T=10.0$ years, rebalanced annually,
from equations (\ref{letf_gbm}-\ref{S_gbm}), GBM case.
LETF leverage factor $\beta = 2.0$.
Numbers in brackets are the standard error estimate at the 95\% confidence level.
Stock data fit of equation (\ref{S_gbm}) to inflation adjusted CRSP index, 1926:1-2023:12.
Interest rate $r$ from T-bills, inflation adjusted, 1926:1-2023:12.
Data in Table \ref{data_table}.   $10^5$ MC simulations.
ES(5\%) is the mean of the worst 5\% of the outcomes.
$\Omega(1)$ defined in equation (\ref{omega_def}).
}
\label{MC_stats_ten}
\end{center}
\end{table}
}

\begin{figure}[htb!]
\centerline{%
\begin{subfigure}[t]{.48\linewidth}
\centering
\includegraphics[width=\linewidth]{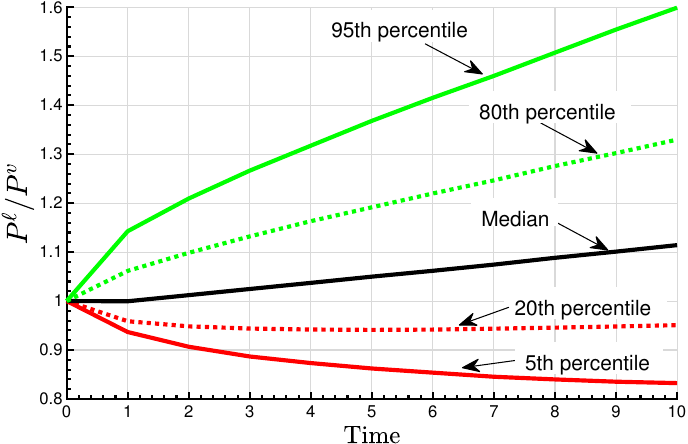}
\caption{Percentiles $P^\ell/P^v$.}
\label{percentile_ratio_gbm_four_five_fig}
\end{subfigure} \qquad
\begin{subfigure}[t]{.48\linewidth}
\centering
\includegraphics[width=\linewidth]{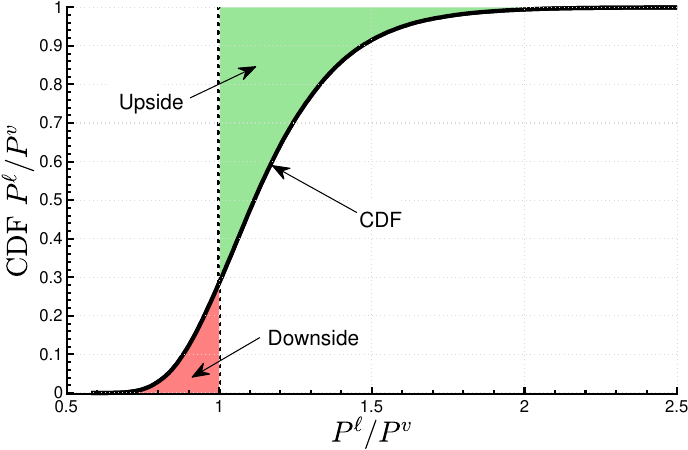}
\caption{CDF $P^{\ell}/P^{v}$. Outperformance if $P^{\ell}/P^{v} >1$. }
\label{CDF_ratio_gbm_four_five_fig}
\end{subfigure}
}
\caption{
Comparing $P^{\ell}/P^v$. $\alpha^\ell = 0.45, \alpha^v = 0.60$. $10^5$ MC simulations.  $T=10.0$ years,
from equations (\ref{letf_gbm}-\ref{S_gbm}), GBM case.
Annual rebalancing.
LETF leverage factor $\beta = 2.0$.
Stock data fit of equation (\ref{S_gbm}) to inflation adjusted CRSP index, 1926:1-2023:12.
Interest rate $r$ from T-bills, inflation adjusted, 1926:1-2023:12.
Data in Table \ref{data_table}.  $T=1$ yr.
$\alpha^{\ell} = 0.45, \alpha^v = 0.6$.
$\Omega(1)$ (equation (\ref{omega_def})) is the ratio of the upside area to the downside area.
}
\label{Ratio_four_five_fig}
\end{figure}

Further insight regarding  the behaviour of the ten year rebalanced LETF 
strategy with $\alpha^\ell = 0.45$ can be seen in Figure \ref{Ratio_four_five_fig}.
The median, 80th and 95th percentiles of $P^{\ell}/P^{v}$ are monotonically increasing
over time,
and greater than one. The 20th percentile of $R_t=P^{\ell}/P^{v}$ stabilizes at about
$0.95$, while the 5th percentile is $\simeq 0.83$ at the ten year mark.  We remind
the reader that these are pathwise measures, hence this may be an acceptable
level of risk for enhanced upside.  Note that the LETF portfolio is rebalanced
to $0.55$ in bonds, while the benchmark strategy has only $0.40$ in bonds
at each rebalancing date.
Figure \ref{CDF_ratio_gbm_four_five_fig} shows the CDF of $R_T=P^{\ell}(T)/P^{v}(T)$, which
illustrates the favourable Omega ratio with the $\alpha^\ell = 0.45$ strategy.

\clearpage

\section{Jump Diffusion\label{sec:Jump-Diffusion results}}
We will now assume that the underlying stock index
follows a jump diffusion process (\ref{S_sde_jumps}),
with double exponential jump size distribution (\ref{eq:dist}).
The final expression for the value of
the LETF portfolio is given in equation (\ref{letf_portfolio}),
and the final value of the VETF portfolio is given in equation (\ref{vetf_portfolio}).
Note that the payoff of the LETF portfolio, as a function of the return
of the underlying stock index $\left( S(T)/S(0) \right)$ is no longer
deterministic, in contrast to the GBM case.
The SDE parameters are fit to the CRSP data 1926:1-2023:12,
see Appendix \ref{jump_params}.

Table \ref{MC_stats_ten_jumps} shows the results with $\alpha^\ell = 0.45$ using various rebalancing
frequencies, along with the case of $\alpha^\ell = 0.30$ (rebalancing annually).  In all cases,
the VETF portfolio is rebalanced to a weight of $\alpha^v= 0.60$ in stocks.

As for the GBM case,  the result using $\alpha^\ell = 0.30$ is unimpressive.
In particular, the mean and median of $R_T = P^{\ell}(T)/P^v(T)$ are less than one, 
and $\Omega(1) = 0.47$ indicating more downside than upside.

In contrast, the results for $\alpha^\ell = 0.45$ are more interesting.  For all rebalancing
frequencies, the mean and median of $R_T$ are larger than one, and $\Omega(1)  > 5.0 $.
Of course, this comes at the cost of more left tail risk, as seen in
the ES(5\%) and the 5th percentile.
As the rebalancing interval decreases, the 95th percentile of $R_T$ decreases slightly,
balanced by an increase in the 5th percentile.  $\Omega(1)$ also increases as
the rebalancing interval decreases.

\begin{figure}[htb!]
\centerline{%
\begin{subfigure}[t]{.48\linewidth}
\centering
\includegraphics[width=\linewidth]{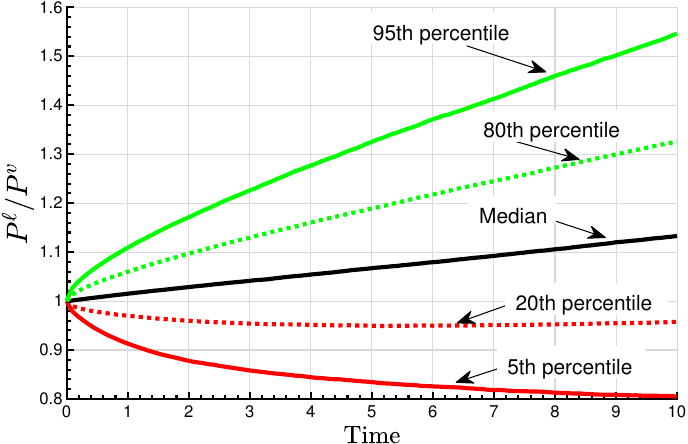}
\caption{Percentiles $P^\ell/P^v$. 20th percentile at $t=10$ is $0.96$.
80th percentile (t=10): $1.33$.}
\label{percentile_ratio_jump_four_five_fig}
\end{subfigure}\qquad	
\begin{subfigure}[t]{.48\linewidth}
\centering
\includegraphics[width=\linewidth]{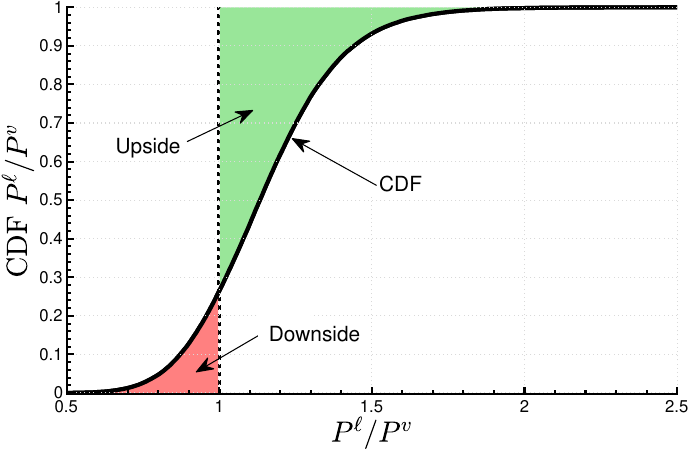}
\caption{CDF $P^{\ell}/P^{v}$. Outperformance if $P^{\ell}/P^{v} >1$. }
\label{CDF_ratio_jump_four_five_fig}
\end{subfigure}
}
\caption{
Comparing $P^{\ell}/P^v$. $\alpha^\ell = 0.45, \alpha^v = 0.60$. $10^5$ MC simulations.  $T=10.0$ years.
Jump diffusion case,
see equations (\ref{letf_portfolio}) and (\ref{vetf_portfolio}).
Monthly rebalancing.
LETF leverage factor $\beta = 2.0$.
SDE parameters  fit  to inflation adjusted CRSP index, 1926:1-2023:12.
See Appendix \ref{jump_params} for the parameters.
$\alpha^{\ell} = 0.45, \alpha^v = 0.6$.
$\Omega(1)$ (equation (\ref{omega_def})) is the ratio of the upside area to the downside area.
}
\label{Ratio_four_five_fig_jump}
\end{figure}

Figure \ref{Ratio_four_five_fig_jump} gives more details for the simulation
with $\alpha^\ell = 0.45$ and monthly rebalancing.  The median, 80th and 95th percentiles
of $R(t) = P^{\ell}(t)/P^v(t)$ increase monotonically from one, and the 20th percentile
stabilizes to end up with a value of $[R_T]_{20th} = 0.96$.  The risk does show up at the
5th percentile (0.8 at ten years).  However, this level of risk may be quite
acceptable, given the Omega(1) ratio, as seen in Figure \ref{CDF_ratio_jump_four_five_fig}.

{\small
\begin{table}[hbt!]
\begin{center}
\begin{tabular}{lccccccc} \toprule[1pt]
Rebalancing   & E[$R_T$] &  Median[$R_T$] & $R_T: 5^{th}$  & $R_T: 95^{th}$   & ES( 5\% )& $\Omega(1)$ &$Prob[R_T > 1]$\\
 interval     &          &                &percentile      & percentile       &          &             & \\
 \hline
          & \multicolumn{7}{c}{$\alpha^\ell = 0.30$} \\
 \hline
Yearly        & 0.9831 (.0004)& 0.9769   & 0.9026         & 1.0838           & 0.8671   & 0.4690     & 0.3139 \\
\hline
         & \multicolumn{7}{c}{$\alpha^\ell = 0.45$} \\
\hline
Yearly       & 1.1561 (.002) & 1.1228     & 0.7712         & 1.6502          & 0.6952    & 5.0141 & 0.7012 \\
Quarterly    &  1.1504 (.001) & 1.1313    & 0.7983         & 1.5646          & 0.7239    & 5.6502 & 0.7299 \\
Monthly      & 1.1494 (.001)  & 1.1330     & 0.8033        & 1.5487          & 0.7308    & 5.8412 & 0.7367 \\
\bottomrule[1pt]
\end{tabular}
\caption{
Statistics for $R(T) = P^{\ell}(T)/P^v(T)$, $10^5$ MC simulations. $\alpha^v = 0.6$.
$T=10.0$ years, rebalancing intervals shown.
Jump diffusion case,
see equations (\ref{letf_portfolio}) and (\ref{vetf_portfolio}).
LETF leverage factor $\beta = 2.0$.
Numbers in brackets are the standard error estimate at the 95\% confidence level.
SDE parameters  fit  to inflation adjusted CRSP index, 1926:1-2023:12.
See Appendix \ref{jump_params} for the parameters.
ES(5\%) is the mean of the worst 5\% of the outcomes.
}
\label{MC_stats_ten_jumps}
\end{center}
\end{table}
}

\section{Summary of Fixed Weight Simulations \label{sec:Fixed weight simulations}}
The overall trends observed for  the results obtained by rebalancing to fixed weights
are consistent for both GBM and jump diffusion models.

At first sight, a reasonable strategy might appear to choose the LETF weight as
$\alpha^\ell = 0.3$, with  $\alpha^v = 0.6$.  This would give both portfolios similar
exposure to the stock index,\footnote{Recall that $\beta = 2.0$.} with the borrowing cost of the LETF offset
by the larger bond component of $P^\ell$.  The idea here would be to gain exposure
to the nonlinear left and right tails of the payoff (see Figure \ref{payoff_fig_point_three}).
However, for reasonable market parameters, this advantage is outweighed
by the LETF fee and the volatility/jump drag, see Remark \ref{drag_remark}.

Increasing the exposure of the LETF with a weight $\alpha^\ell = 0.45$, with
$\alpha^v = 0.60$, results in a much more interesting strategy.  In this case,
there is a gain in median, 80th and 95th percentile's of $R_T = P^\ell(T)/P^{v}(T)$,
which is offset by an increased left tail risk.  However, some investors
may find this tradeoff appealing.

The key feature which contrasts the $\alpha^\ell = 0.3$ with the case $\alpha^\ell = 0.45$
appears to be the effect the Omega ratio.  For one year investment horizons,
the Omega ratio for $\alpha^\ell = 0.3$ is less than one, while the Omega
ratio for $\alpha^\ell = 0.45$ is larger than one.  This effect seems to compound
for more rebalancing times over longer intervals, in the sense that the Omega
ratio for $\alpha^\ell = 0.3$ decreases for longer time horizons, while
the Omega ratio for $\alpha^\ell = 0.45$ increases for longer time horizons.

Of course, we have used a very simple strategy, involving constant weight
rebalancing.  Different choices of $\alpha^\ell$ will result in different
tradeoffs between risk and reward.  However, these results indicate that
LETFs can be advantageous even with fairly naive strategies.

Clearly, dynamic allocation of the amount invested in the LETF should result
in even better investment policies.  This requires use of more sophisticated
algorithms for asset allocation.  In addition, use of parametric SDE models
of asset price dynamics is prone to misspecification.

In the next Section, we will explore the use of a Neural Network
asset allocation strategy, coupled with a data-driven approach
to simulating portfolios containing LETFs and VETFs.

\section{Data-driven Machine Learning Approach\label{sec:Data-driven-Machine-Learning}}

Instead of using fixed portfolio allocations to the LETF as in Section
\ref{sec:Fixed weight simulations}, we now use a data-driven neural
network approach to determine the optimal \textit{dynamic} allocation
of wealth to the LETF at each rebalancing time. The proportion to
invest in the LETF is given by a neural network, and we assume that
there is no short-selling or leveraging of the investment in the LETF.

The neural network is trained on nearly a century of bootstrapped
historical data, spanning 1926:01 to 2023:12, where LETF and VETF
returns are synthetically constructed. Learning the optimal investment
strategy using such a long data period ensures that periods of exceptional
market volatility and high inflation are also included, rather than
focusing on the most recent decades since the inception of LETFs.
See Appendix \ref{sec: Appendix Constructing-synthetic-LETF} for
more information on the construction of proxy returns time series
for the LETF and VETF.  We remind the reader that all returns
are inflation adjusted.

We emphasize the use of stationary block bootstrap resampling rather
than fixed block bootstrap resampling to generate training/testing
data for the neural network. In summary, the stationary block bootstrap
method (\citet{politis1994}) randomly varies block lengths according
to a geometric distribution, which better preserves the temporal dependence
structure of financial returns while introducing greater variability
in the resampled sequences. This approach addresses two key limitations
of fixed block bootstrap; first, it reduces the repetitiveness that
can arise from using uniform block lengths, and second, it maintains
adequate variance in the resampled return series by avoiding overly
rigid partitioning of the original time series. Furthermore, the stationary
block bootstrap method is popular both in in academic settings (\citet{Cederburg_2022})
and practical applications (\citet{CogneauZakalmouline2013,dichtl2016,Scott_2017,Scott_2022,Simonian_2022}).

We assume that both the LETF and VETF portfolios are rebalanced quarterly,
i.e. at $N_{rb}=40$ discrete rebalancing times during the investment
time horizon $\left[t_{0}=0,T=10\textrm{ years}\right]$,

\begin{eqnarray}
\mathcal{\mathcal{T}} & = & \left\{ \left.t_{n}=n\Delta t\right|n=0,...,39\right\} ,\qquad\Delta t=0.25.\label{eq: Set of rebal times}
\end{eqnarray}
In contrast to the previous sections, we now do not need to specify
any parametric dynamics for the underlying assets. Instead, the neural
network will simply learn the optimal investment in the LETF using
bootstrapped historical returns, which incorporates all empirical
return characteristics, including for example fat tails, volatility
clustering, serial correlation, asymmetric return distributions, stochastic
volatility, inflation regime changes, and other stylized facts of
financial markets that are difficult to capture in parametric models.
Suppose the bootstrapped returns over the time interval $\left[t_{n-1},t_{n}\right]$
for the LETF, VETF and 30-day T-bill are given by $R_{\ell}\left(t_{n}\right),R_{v}\left(t_{n}\right)$
and $R_{B}\left(t_{n}\right)$ respectively. Assuming zero contributions,
the LETF portfolio $P^{\ell}$ and VETF portfolio $P^{v}$ have the
following dynamics respectively, 
\begin{eqnarray}
P^{\ell}\left(t_{n+1}\right) & = & P^{\ell}\left(t_{n}\right)\cdot\left[\alpha^{\ell}\left(t_{n-1}\right)\cdot\left(1+R_{\ell}\left(t_{n}\right)\right)+\left(1-\alpha^{\ell}\left(t_{n-1}\right)\right)\cdot\left(1+R_{B}\left(t_{n}\right)\right)\right],\label{eq: LETF portfolio value dynamics}\\
P^{v}\left(t_{n+1}\right) & = & W\left(t_{n}\right)\cdot\left[\alpha^{v}\cdot\left(1+R_{v}\left(t_{n}\right)\right)+\left(1-\alpha^{v}\right)\cdot\left(1+R_{B}\left(t_{n}\right)\right)\right].\label{eq: VETF portfolio value dynamics}
\end{eqnarray}

We continue assuming a proportion in the VETF of $\alpha^{v}=0.60$
as in previous sections, but now we wish to determine the dynamic
proportion to invest in the LETF at each rebalancing time, $\alpha^{\ell}\left(t_{n}\right),t_{n}\in\mathcal{\mathcal{T}}.$
What is more, we want to determine an \textit{optimal} proportion
in the LETF, $\alpha^{\ell\ast}\left(t_{n}\right),t_{n}\in\mathcal{\mathcal{T}}$,
which requires the specification of the an investment objective. For this purpose,
we choose the cumulative tracking difference or CD objective (see
\citet{PvSForsythLi2022_CD}) aimed at targeting a favourable tracking
differences of the LETF portfolio relative to the VETF portfolio over
the investment time horizon. With annual outperformance target parameter
$\delta$, we consider the CD objective over a time horizon of $T=10$
years with quarterly rebalancing,
\begin{eqnarray}
\left(CD\left(\delta\right)\right): &  & \inf_{\alpha^{\ell}\left(t_{n}\right),t_{n}\in\mathcal{\mathcal{T}}}\left[\sum_{n=0}^{39}\left(P^{\ell}\left(t_{n}\right)-e^{\delta\cdot t_{n}}\cdot P^{v}\left(t_{n}\right)\right)^{2}\right].\label{eq: CD objective}
\end{eqnarray}

Note that we could formulate an alternative version of \eqref{eq: CD objective}
in terms of the ratio (see \eqref{omega_def}), for example 
\begin{equation}
\inf_{\alpha^{\ell}\left(t_{n}\right),t_{n}\in\mathcal{\mathcal{T}}}\left[\sum_{n=0}^{39}\left(\frac{P^{\ell}\left(t_{n}\right)}{P^{v}\left(t_{n}\right)}-e^{\delta\cdot t_{n}}\right)^{2}\right].\label{eq: CD objective in RATIO form}
\end{equation}
While \eqref{eq: CD objective in RATIO form} is not exactly equivalent
to \eqref{eq: CD objective} in a mathematical sense, the resulting
investment strategies are qualitatively similar\footnote{The can be seen by using either numerical experiments, or assuming
for example GBM dynamics for the underlying assets and obtaining the
closed form solutions for a continuous-time version of \eqref{eq: CD objective in RATIO form}
using techniques as in \citet{PvSForsythLi2022_CD}.}, while \eqref{eq: CD objective} is more convenient in terms of gradient
flows (for neural network training) and in terms of the range of VETF
portfolio strategies that can be incorporated. For illustrative purposes,
we therefore continue working with \eqref{eq: CD objective}. 

How can \eqref{eq: CD objective} be solved without difficulty? We
use the neural network approach of \citet{LiForsyth2019,PvsForsythLi2023_NN}),
which can be classified as a ``global-in-time'' machine learning
approach (see \citet{HuLauriere2023}) to stochastic control problems
like \eqref{eq: CD objective}. Only a single optimization problem
is solved which determines the parameters of a single neural network,
with time and total portfolio values as feature variables.
In addition, since the neural network uses an architecture that enforces
the investment constraints of no short-selling and no leveraged positions
automatically, the optimization in \eqref{eq: CD objective} is unconstrained.
For more information and solution implementation details, see \citet{LiForsyth2019,PvsForsythLi2023_NN}. 

Once the problem \eqref{eq: CD objective} is solved, we can obtain
the optimal dynamic allocation to the LETF $\alpha^{\ell\ast}\left(t_{n}\right)$
at any rebalancing time $t_{n}\in\mathcal{\mathcal{T}}$, as the output
of the neural network for inputs features time $t_{n}$, LETF portfolio
value $P^{\ell}\left(t_{n}\right)$ and VETF portfolio value $P^{v}\left(t_{n}\right)$. 

Considering two values of the annual outperformance target parameter
$\delta$ in the CD objective \eqref{eq: CD objective}, namely $\delta=0.02$
and $\delta=0.04$, we consider the summary statistics in Table \ref{tbl_Num_Stats_CD_obj}
of the terminal ratio $R(T)=P^{\ell}(T)/P^{v}(T)$ obtained using
the corresponding optimal LETF allocations {\small{}$\alpha^{\ell\ast}$}
and VETF allocation of $\alpha^{v}=0.6$. Comparing Table \ref{tbl_Num_Stats_CD_obj}
with the results of previous sections, for example Table \ref{MC_stats_ten_jumps},
we observe a significant improvement in the Omega ratio $\Omega(1)$
and $Prob[R_{T}>1]$ from using the optimal dynamic strategies $\alpha^{\ell\ast}$
rather than a fixed weight allocation $\alpha^{\ell}$ to the LETF.
However, we also note that there is no free lunch, in the sense that
the significant improvement in performance as per the Omega ratio
also accompanied by an increase in downside risk, as observed by a
reduction in the mean of the worst 5\% of outcomes ({\small{}ES( 5\%
)}) of $R(T)$.

\noindent {\small{}}
\begin{table}[hbt!]
\centering{}{\small{}}%
\begin{tabular}{lccccccc}
\toprule 
{\small{}{}Rebalancing} & {\small{}E{[}$R_{T}${]}} & {\small{}Median{[}$R_{T}${]}} & {\small{}$R_{T}:5^{th}$} & {\small{}$R_{T}:95^{th}$} & {\small{}ES( 5\% )} & {\small{}$\Omega(1)$} & {\small{}$Prob[R_{T}>1]$}\tabularnewline
{\small{}interval} &  &  & {\small{}percentile} & {\small{}percentile} &  &  & \tabularnewline
\midrule 
 & \multicolumn{7}{c}{{\small{}Using optimal $\alpha^{\ell\ast}$ for $CD\left(\delta=0.02\right)$}}\tabularnewline
\midrule 
{\small{}Quarterly} & {\small{}1.1153 (<0.001)} & {\small{}1.1423} & {\small{}0.8857} & {\small{}1.2407} & {\small{}0.7134} & {\small{}7.8316} & {\small{}0.8956}\tabularnewline
\midrule 
 & \multicolumn{7}{c}{{\small{}Using optimal $\alpha^{\ell\ast}$ for $CD\left(\delta=0.04\right)$}}\tabularnewline
\midrule 
{\small{}Quarterly} & {\small{}1.2419 (<0.001)} & {\small{}1.3073} & {\small{}0.6995} & {\small{}1.4576} & {\small{}0.4919} & {\small{}8.0831} & {\small{}0.8814}\tabularnewline
\midrule 

\end{tabular}{\small{}\caption{Statistics for $R(T)=P^{\ell}(T)/P^{v}(T)$, $5\times10^{5}$ bootstrapped
historical data paths. Optimal LETF allocation {\small{}$\alpha^{\ell\ast}$}
determined using CD objective function with targets $\delta$ as shown,
with VETF allocation of $\alpha^{v}=0.6$. $T=10.0$ years, quarterly
rebalancing. LETF leverage factor $\beta=2.0$. Numbers in brackets
are the standard error estimate of the mean at the 95\% confidence
level. ES(5\%) is the mean of the worst 5\% of the outcomes.}
\label{tbl_Num_Stats_CD_obj}}
\end{table}
{\small\par}

\begin{figure}[htb!]
\centerline{\begin{subfigure}[t]{.48\linewidth} \centering \includegraphics[width=1\linewidth]{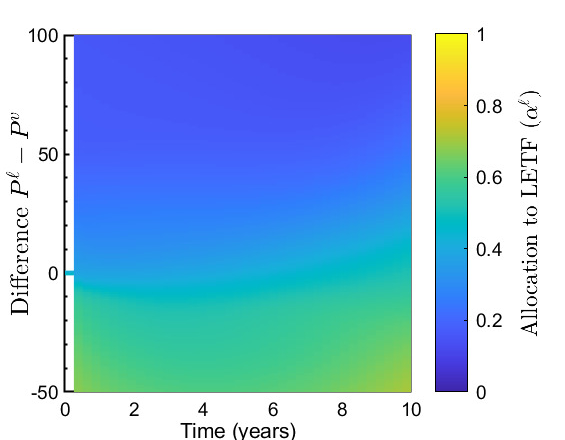}
\caption{Optimal $\alpha^{\ell\ast}$ for $CD\left(\delta=0.02\right)$.}
\label{fig_Num_Heatmaps_outperf_02} \end{subfigure}\qquad{}\begin{subfigure}[t]{.48\linewidth}
\centering \includegraphics[width=1\linewidth]{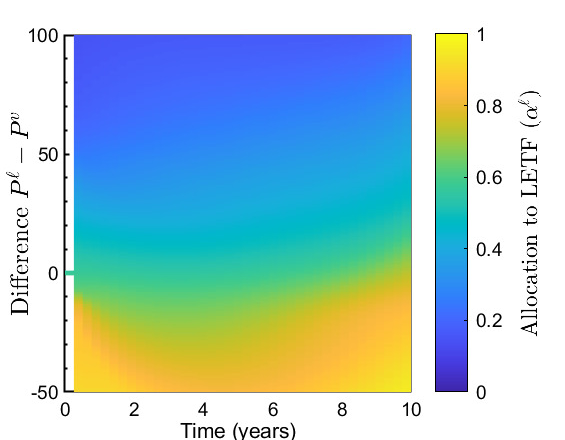}
\caption{Optimal $\alpha^{\ell\ast}$ for $CD\left(\delta=0.04\right)$. }
\label{fig_Num_Heatmaps_outperf_04} \end{subfigure} } \caption{Optimal dynamic LETF allocation $\alpha^{\ell\ast}\left(t_{n}\right)$
illustrated as a function of the difference $P^{\ell}\left(t_{n}\right)-P^{v}\left(t_{n}\right)$
and time $t_{n}$ (years), using the CD objective function with for
two different values of the annual outperformance target parameter
$\delta$. At time $t=0$, the difference is $P^{\ell}\left(t_{0}\right)-P^{v}\left(t_{0}\right)=0$
by definition. The LETF leverage factor is $\beta=2.0$. }
\label{fig_Num_Heatmaps} 
\end{figure}

To compare the optimal dynamic strategies $\alpha^{\ell\ast}$ for
$CD\left(\delta=0.02\right)$ and $CD\left(\delta=0.04\right)$ to
the fixed allocations $\alpha^{\ell}$ to the LETF considered in the
preceding sections, Figure \ref{fig_Num_Heatmaps} illustrates $\alpha^{\ell\ast}\left(t_{n}\right)$
as a function of the difference $P^{\ell}\left(t_{n}\right)-P^{v}\left(t_{n}\right)$
and time $t_{n}$. As expected given the outperformance targets, the
optimal allocation $\alpha^{\ell\ast}$ for $CD\left(\delta=0.02\right)$
illustrated in Figure \ref{fig_Num_Heatmaps_outperf_02} is generally
smaller (or less aggressive) than the corresponding allocation $\alpha^{\ell\ast}$
for $CD\left(\delta=0.04\right)$ illustrated in Figure \ref{fig_Num_Heatmaps_outperf_02}.
However, both Figure \ref{fig_Num_Heatmaps_outperf_02} and Figure
\ref{fig_Num_Heatmaps_outperf_02} illustrates how the optimal LETF
strategy $\alpha^{\ell\ast}$ is contrarian, in the sense that smaller
(resp. larger) values of the difference $P^{\ell}\left(t_{n}\right)-P^{v}\left(t_{n}\right)$
results in larger (resp. smaller) allocations to the LETF. In other
words, following strong (resp. weak) LETF returns, the investment
in the LETF is decreased (resp. increased). This enables the investor
to ``lock-in'' periods of strong returns while simultaneously reducing
risk.

\noindent 

\begin{figure}[htb!]
\centerline{\begin{subfigure}[t]{.48\linewidth} \centering \includegraphics[width=1\linewidth]{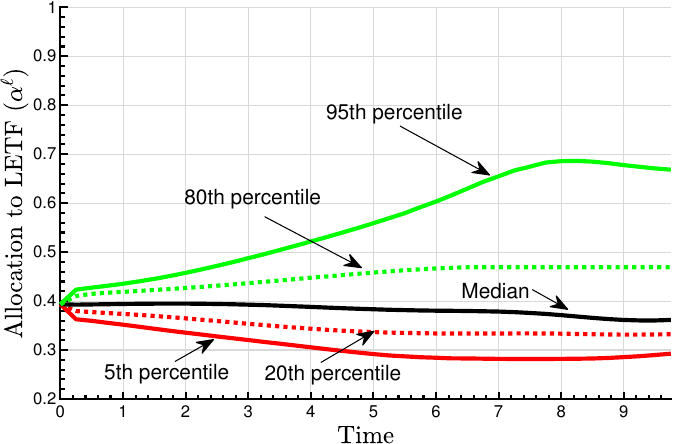}
\caption{Percentiles of optimal $\alpha^{\ell\ast}$ for $CD(\delta=0.02)$.}
\label{fig_Num_pctiles_LETF_alloc_outperf_02} \end{subfigure}\qquad{}\begin{subfigure}[t]{.48\linewidth}
\centering \includegraphics[width=1\linewidth]{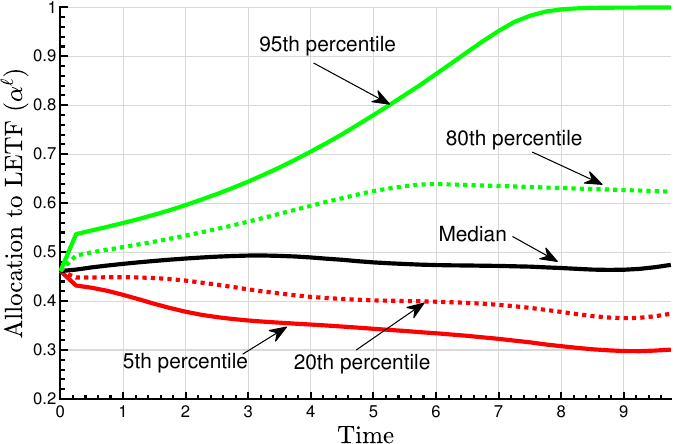}
\caption{Percentiles of optimal $\alpha^{\ell\ast}$ for $CD(\delta=0.04)$.}
\label{fig_Num_pctiles_LETF_alloc_outperf_04} \end{subfigure} }
\caption{Percentiles of the optimal dynamic LETF allocation $\alpha^{\ell\ast}\left(t_{n}\right)$
over time using the CD objective function with two different values
of the annual outperformance target parameter $\delta$. Results shown
using $5\times10^{5}$ bootstrapped historical data paths. LETF leverage
factor $\beta=2.0$. The same scale on the $y$-axis is used to facilitate
comparison. }
\label{fig_Num_pctiles_LETF_alloc} 
\end{figure}

Of course, not all values of $\alpha^{\ell\ast}$ illustrated in Figure
\ref{fig_Num_Heatmaps} are equally likely to be implemented. Figure
\ref{fig_Num_pctiles_LETF_alloc} illustrates selected percentiles
of the optimal allocation $\alpha^{\ell\ast}$ over time on the bootstrapped
historical data used to train the neural network. We observe that
in the case of $\alpha^{\ell\ast}$ for the $CD(\delta=0.02)$ objective
(Figure \ref{fig_Num_pctiles_LETF_alloc_outperf_02}), it is typical
for $\alpha^{\ell\ast}$ to take on values in the interval $\left[0.30,0.45\right]$,
with median values of around 0.35 to 0.4. In the case of $\alpha^{\ell\ast}$
for the $CD(\delta=0.04)$ objective (Figure \ref{fig_Num_pctiles_LETF_alloc_outperf_04}),
while $\alpha^{\ell\ast}$ can take on more extreme values due to
the more aggressive outperformance target, the median value is around
0.5. These results are perhaps expected given the the simulations
of the preceding sections: to achieve the high Omega ratio results
illustrated in Table \ref{tbl_Num_Stats_CD_obj} over the longer time
horizon of 10 years, the optimal dynamic allocation strategy $\alpha^{\ell\ast}$
reduces exposure to the LETF to below 0.3 slowly (see 5th percentiles
in Figure \ref{fig_Num_pctiles_LETF_alloc}), and then only when the
LETF has performed well compared to the VETF in prior periods (see
Figure \ref{fig_Num_Heatmaps}). 

\noindent 

\begin{figure}[htb!]
\centerline{\begin{subfigure}[t]{.48\linewidth} \centering \includegraphics[width=1\linewidth]{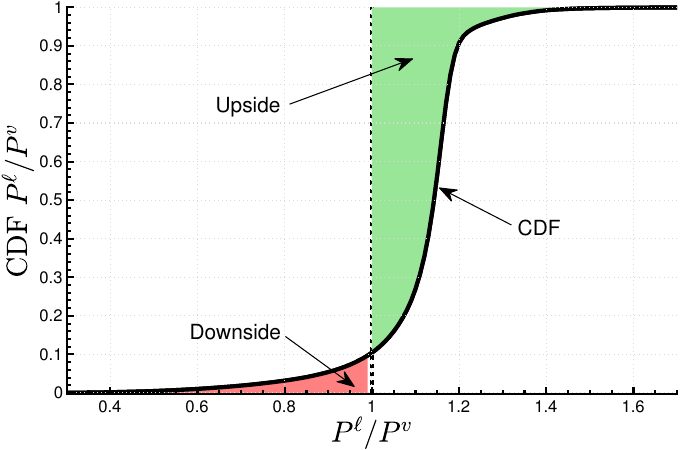}
\caption{CDF $P^{\ell}/P^{v}$ using optimal $\alpha^{\ell\ast}$ for $CD(\delta=0.02)$.}
\label{fig_Num_CDF_ratio_outperf_02} \end{subfigure}\qquad{}\begin{subfigure}[t]{.48\linewidth}
\centering \includegraphics[width=1\linewidth]{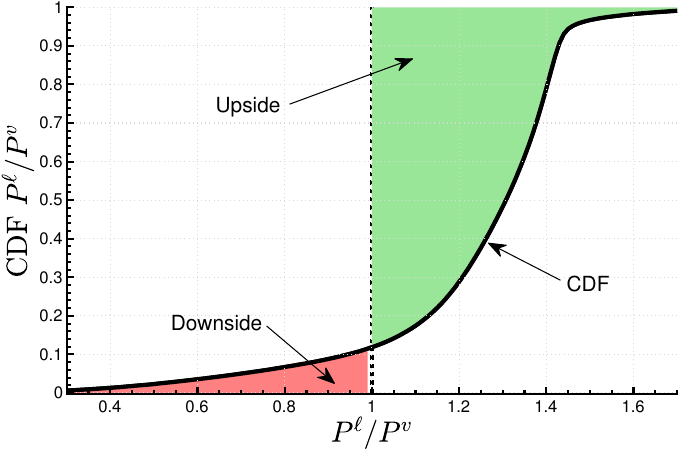}
\caption{CDF $P^{\ell}/P^{v}$ using optimal $\alpha^{\ell\ast}$ for $CD(\delta=0.04)$. }
\label{fig_Num_CDF_ratio_outperf_04} \end{subfigure} } \caption{ Comparing $P^{\ell}/P^{v}$ at time $T=10.0$ years, outperformance
if $P^{\ell}/P^{v}>1$. Optimal dynamic $\alpha^{\ell\ast}$ is used,
obtained via the CD objective function with two different values of
the annual outperformance target parameter $\delta$, $\alpha^{v}=0.60$.
Results shown using $5\times10^{5}$ bootstrapped historical data
paths. LETF leverage factor $\beta=2.0$. $\Omega(1)$ (equation (\ref{omega_def}))
is the ratio of the upside area to the downside area. }
\label{fig_Num_CDF_ratio} 
\end{figure}

Recalling from Table \ref{tbl_Num_Stats_CD_obj} that we obtain $\Omega(1)=7.83$
when using $\alpha^{\ell\ast}$ for $CD(\delta=0.02)$, whereas the
Omega ratio increases to $\Omega(1)=8.08$ when using $\alpha^{\ell\ast}$
for $CD(\delta=0.04)$. In other words, the more aggressive outperformance
target of $\delta=0.04$ in the CD objective also increased the Omega
ratio of the resulting strategy. However, Figure \ref{fig_Num_CDF_ratio}
illustrates what can already be deduced from Table \ref{tbl_Num_Stats_CD_obj},
namely that as the target increases from $\delta=0.02$ to $\delta=0.04$,
both the upside and the the downside outcomes increase in terms of
frequency and severity. However, with the higher target $\delta=0.04$
the upside increases significantly \textit{more} relative to the increase
in the downside, resulting in the higher observed Omega ratio in Table
\ref{tbl_Num_Stats_CD_obj}. 

\noindent 

\begin{figure}[htb!]
\centerline{\begin{subfigure}[t]{.48\linewidth} \centering \includegraphics[width=1\linewidth]{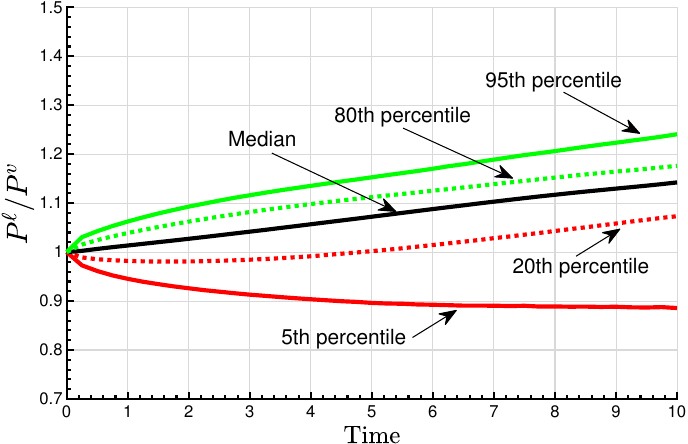}
\caption{Percentiles of $P^{\ell}/P^{v}$ using optimal $\alpha^{\ell\ast}$
for $CD(\delta=0.02)$.}
\label{fig_Num_pctiles_ratio_outperf_02} \end{subfigure}\qquad{}\begin{subfigure}[t]{.48\linewidth}
\centering \includegraphics[width=1\linewidth]{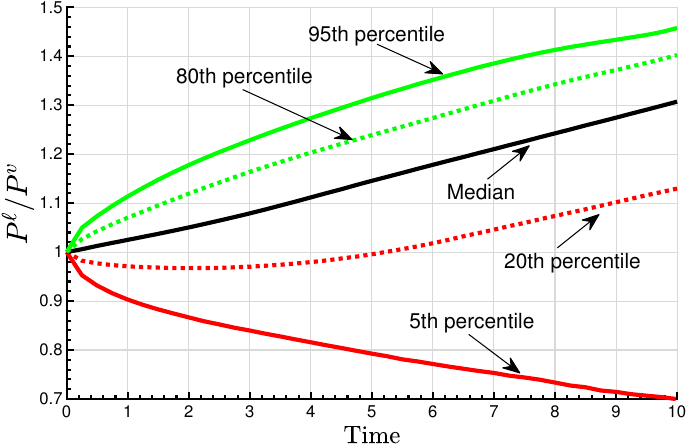}
\caption{Percentiles of $P^{\ell}/P^{v}$ using optimal $\alpha^{\ell\ast}$
for $CD(\delta=0.04)$. }
\label{fig_Num_pctiles_ratio_outperf_04} \end{subfigure} } \caption{ Percentiles of $P^{\ell}/P^{v}$ over time, outperformance if $P^{\ell}/P^{v}>1$.
Optimal dynamic $\alpha^{\ell\ast}$ is used, obtained via the CD
objective function with two different values of the annual outperformance
target parameter $\delta$, $\alpha^{v}=0.60$. Results shown using
$5\times10^{5}$ bootstrapped historical data paths. LETF leverage
factor $\beta=2.0$. The same scale on the $y$-axis is used to facilitate
comparison. }
\label{fig_Num_pctiles_ratio} 
\end{figure}

Figure \ref{fig_Num_pctiles_ratio} considers the percentiles of the
portfolio value ratio $P^{\ell}/P^{v}$ over time when using the optimal
dynamic LETF allocation $\alpha^{\ell\ast}$. Note that the same scale
on the $y$-axis is used to facilitate comparison. We observe that
the the optimal $\alpha^{\ell\ast}$ for $CD(\delta=0.04)$ outperforms
the optimal $\alpha^{\ell\ast}$ for $CD(\delta=0.02)$ in terms of
all percentiles shown for the ratio $P^{\ell}/P^{v}$ except for the
5th percentile, which is significantly worse using the more aggressive
target. However, as noted in Section \ref{sec:Fixed weight simulations},
this might be an appealing tradeoff for some investors.

\noindent 

Since future asset returns will never replicate historical return
paths precisely, we consider the preceding illustrative investment
results using bootstrapped historical data (i.e. using stationary
block bootstrap resampling) to be significantly more informative than
single historical return path subsets when illustrating performance. 

However, some practitioners may find bootstrapped results abstract
or difficult to interpret intuitively. To provide more relatable and
concrete illustrations, we examine how the optimal dynamic $\alpha^{\ell\ast}$
LETF portfolios and benchmark VETF portfolio would have performed
during distinct 10-year historical periods. Table \ref{tbl_Num_Historical_W_Ts}
demonstrates the portfolio performance starting with \$100 initial
wealth invested over $T=10$ years, beginning at the indicated month.
Note that these results necessarily also rely on synthetically constructed
historical returns for LETFs (see Appendix \ref{sec: Appendix Constructing-synthetic-LETF}).
Table \ref{tbl_Num_Historical_W_Ts} confirms the advantages of strategically
incorporating LETFs into investment portfolios for long-term investors,
even if only using infrequent (quarterly) rebalancing.

\noindent {\small{}}
\begin{table}[hbt!]
\noindent \begin{centering}
\begin{tabular}{|>{\centering}p{2cm}|>{\centering}p{2cm}|>{\centering}p{3.2cm}|>{\centering}p{3.2cm}|>{\centering}p{3.2cm}|}
\hline 
\multirow{3}{2cm}{{\small{}Starting month of investment:}} & \multirow{3}{2cm}{{\small{}Final month ($T=10$ years)}} & \multicolumn{3}{>{\centering}p{9.6cm}|}{{\small{}Terminal wealth for each portfolio, initial investment \$100, }{\small\par}

{\small{}quarterly rebalancing, zero contributions}}\tabularnewline
\cline{3-5} \cline{4-5} \cline{5-5} 
 &  & {\small{}VETF portfolio} & {\small{}LETF portfolio} & {\small{}LETF portfolio}\tabularnewline
 &  & {\small{}$\alpha^{v}=0.60$} & {\small{}Dynamic $\alpha^{\ell\ast}$}

{\small{}$CD(\delta=0.02)$} & {\small{}Dynamic $\alpha^{\ell\ast}$}{\small\par}

{\small{}$CD(\delta=0.04)$}\tabularnewline
\hline 
{\small{}Jan 1970} & {\small{}Dec 1979} & {\small{}93} & {\small{}102} & {\small{}108}\tabularnewline
\hline 
{\small{}Jan 1975} & {\small{}Dec 1984} & {\small{}176} & {\small{}200} & {\small{}224}\tabularnewline
\hline 
{\small{}Jan 1980} & {\small{}Dec 1989} & {\small{}225} & {\small{}259} & {\small{}303}\tabularnewline
\hline 
{\small{}Jan 1985 } & {\small{}Dec 1994} & {\small{}197} & {\small{}222} & {\small{}252}\tabularnewline
\hline 
{\small{}Jan 1990} & {\small{}Dec 1999} & {\small{}250} & {\small{}297} & {\small{}359}\tabularnewline
\hline 
{\small{}Jan 1995} & {\small{}Dec 2004} & {\small{}187} & {\small{}212} & {\small{}236}\tabularnewline
\hline 
{\small{}Jan 2000} & {\small{}Dec 2009} & {\small{}88} & {\small{}73} & {\small{}57}\tabularnewline
\hline 
{\small{}Jan 2005} & {\small{}Dec 2014} & {\small{}141} & {\small{}161} & {\small{}186}\tabularnewline
\hline 
{\small{}Jan 2010} & {\small{}Dec 2019} & {\small{}188} & {\small{}218} & {\small{}260}\tabularnewline
\hline 
{\small{}Jan 2013} & {\small{}Dec 2022} & {\small{}164} & {\small{}189} & {\small{}219}\tabularnewline
\hline 
{\small{}Jan 2014} & {\small{}Dec 2023} & {\small{}158} & {\small{}182} & {\small{}216}\tabularnewline
\hline 
\end{tabular}
\par\end{centering}
\centering{}{\small{}\caption{Historical path results: Investing initial wealth of \$100 over $T=10.0$
years, starting at the beginning of the month as indicated, and following
the optimal LETF allocation {\small{}$\alpha^{\ell\ast}$} determined
using CD objective function with targets $\delta$ as shown, as well
as the VETF allocation of $\alpha^{v}=0.6$. Quarterly rebalancing,
LETF leverage factor $\beta=2.0$, zero contributions are made to
the portfolio. }
\label{tbl_Num_Historical_W_Ts}}{\small\par}
\end{table}
{\small\par}

\begin{figure}[htb!]
\centerline{\begin{subfigure}[t]{.48\linewidth} \centering \includegraphics[width=1\linewidth]{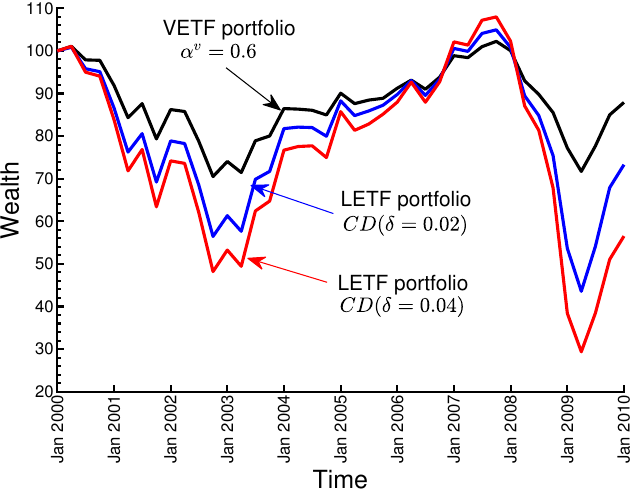}
\caption{Start investing in Jan 2000.}
\label{fig_Num_hist_paths_start_Jan_2000} \end{subfigure}\qquad{}\begin{subfigure}[t]{.48\linewidth}
\centering \includegraphics[width=1\linewidth]{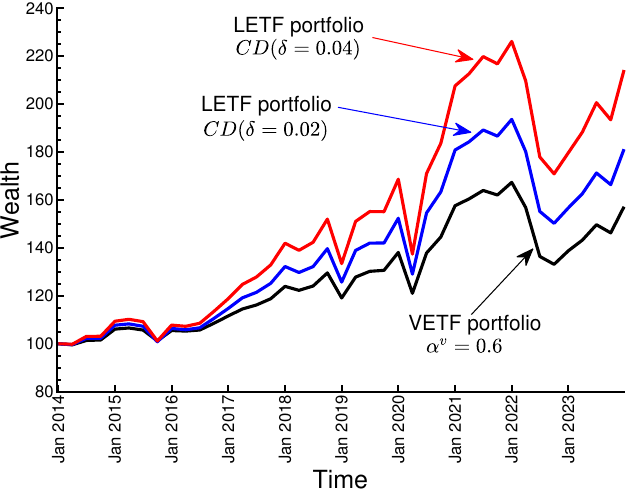}
\caption{Start investing in Jan 2014. }
\label{fig_Num_hist_paths_start_Jan_2014} \end{subfigure} } \caption{Historical path results over time, selected months: Investing initial
wealth of \$100 over $T=10.0$ years, starting at the beginning of
the month as indicated, and following the optimal LETF allocation
{\small{}$\alpha^{\ell\ast}$} determined using CD objective function
with targets $\delta$ as shown, as well as the VETF allocation of
$\alpha^{v}=0.6$. Quarterly rebalancing, LETF leverage factor $\beta=2.0$,
zero contributions are made to the portfolio. }
\label{fig_Num_hist_paths} 
\end{figure}

A key exception to the excellent performance of the LETF portfolios
relative to the VETF portfolio is the period January 2000 to December 2009,
where the VETF portfolio comfortably outperforms both LETF portfolios.
However, this period is special, not only in the sense that it contains
two major stock market events (the dot-com crash and the Global Financial
Crisis), but that the GFC occurs near the end of the investment time
horizon, which disproportionately affects the LETF portfolios. To
illustrate this observation, Figure \ref{fig_Num_hist_paths} compares
the evolution of portfolio wealth over time for the different strategies,
starting in January 2000 and January 2014. Note that the hypothetical
investor starting in January 2014 will experience, over the time horizon
of $T=10$ years, strong early post-GFC growth, followed by the Covid-19
disruptions, subsequent recovery, as well as the 2022 bear market
and a period of rising interest rates and inflation. For both Figure
\ref{fig_Num_hist_paths_start_Jan_2000} and Figure \ref{fig_Num_hist_paths_start_Jan_2014},
we observe that regardless of the stock market crash in question,
the LETF portfolios experience larger peak-to-trough declines but
also faster post-crash recovery. The 10-year period starting in January 2000 therefore terminates in December 2009 before the LETF investor can take advantage of the post-crash recovery observed in Figure \ref{fig_Num_hist_paths_start_Jan_2014}.

\section{Conclusion\label{sec:Conclusion}}

This paper explores how broad stock market LETFs can enhance portfolio
performance when incorporated within dynamic, actively managed strategies,
despite their well-documented unsuitability for passive investment
strategies. 

The results, which are based on synthetic LETF returns constructed
with nearly 100 years of market data, reveals that the key to successful
incorporation of a broad stock market LETF into a portfolio requires
a contrarian investment strategy, i.e. reducing exposure to the LETF
systematically following a period of gains. We observed that the Omega
ratio is a critical metric for evaluating LETF strategies, with results
showing that while naive approaches often yield Omega ratios below
unity, dynamic strategies can achieve substantially higher Omega ratios
that benefits from (infrequent) portfolio rebalancing. Using a data-driven
neural network approach, we discuss the construction of optimal LETF
allocation strategies without restrictive parametric assumptions,
learning the strategy directly from historical returns to capture
market complexity while maintaining practical implementability through
infrequent rebalancing. 

Our results demonstrate that accessible dynamic approaches to broad
stock market LETF investment, requiring only quarterly rebalancing,
can deliver potentially desirable risk-return benefits for the active
investor. 

\section{Declaration}
This work was supported by the Natural Sciences and Engineering Research
Council of Canada (NSERC) under grants RGPIN-2017-03760 and RGPIN-2020-04331.
The authors have no conflicts of interest to report.

\appendix
\section*{Appendices}

\section{Constructing synthetic LETF and VETF returns since 1926\label{sec: Appendix Constructing-synthetic-LETF}}

Since LETFs were only introduced in 2006 (\citet{BansalMarshall2015}),
our goal of deriving a more realistic historical perspective of LETF
behaviour during different market and inflation regimes requires the
use of proxy data. Note that a similar construction of proxy returns
time series for LETFs has also been done in \citet{BansalMarshall2015},
but with details regarding the inflation adjustment and choice of
underlying index differ. As in for example \citet{BansalMarshall2015}
and \citet{LeungSircar2015}, we assume that the ETF managers achieve
a negligible tracking error with respect to the underlying index. 

The steps followed in constructing proxy time series of historical
returns of a VETF and LETF referencing a broad US equity market index,
we proceed as follows:
\begin{enumerate}
\item We obtain daily returns for the VWD index and 30-day Treasury bills
from the CRSP\footnote{Calculations were based on data from the Historical Indexes 2024©,
Center for Research in Security Prices (CRSP), The University of Chicago
Booth School of Business. Wharton Research Data Services was used
in preparing this article. This service and the data available thereon
constitute valuable intellectual property and trade secrets of WRDS
and/or its third party suppliers.}. The CRSP VWD index, capitalization-weighted index consisting of
all domestic stocks trading on major US exchanges, is assumed to be
the broad stock market index referenced by the VETF and LETF. We use
data for the historical time period 1926:01 to 2023:12, to ensure
that periods of exceptional market volatility and high inflation are
also included. 
\item Multiply each daily return by the returns multiplier $\beta$, where
we used $\beta=2$ for the LETF and $\beta=1$ for the VETF, and construct
a time series of monthly returns. This follows from the results in
Section \ref{jump_section}, with $\left(\delta V^{\ell}/V^{\ell}\right)$
and $\left(\delta V^{v}/V^{v}\right)$ representing the daily returns
of the LETF and VETF, respectively:
\begin{eqnarray}
\frac{\delta V^{\ell}}{V^{\ell}} & \simeq & \beta\cdot\left(\frac{\delta S}{S}\right)+\left(1-\beta\right)\cdot\left(\frac{\delta B}{B}\right)-c_{\ell}\cdot\delta t,\label{eq: LETF proxy return time series}\\
\frac{\delta V^{v}}{V^{v}} & \simeq & \frac{\delta S}{S}-c_{v}\cdot\delta t.\label{eq: VETF proxy returns time series}
\end{eqnarray}
In this approximation, $\delta S/S$ denotes the daily returns of
the equity market index, $\delta B/B$ the daily returns of the 30-day
T-bills, and $\delta t=1/252$. To reflect values similar to those
observed in the market (see Section \ref{jump_section}), we assume
expense ratios of $c_{\ell}=.0089$ and $c_{v}=0.00$, and a LETF
leverage factor of $\beta=2$.
\item Finally, all time series were inflation-adjusted using inflation data
from the US Bureau of Labor Statistics\footnote{The annual average CPI-U index, which is based on inflation data for
urban consumers, were used - see \texttt{http://www.bls.gov.cpi}}.
\end{enumerate}
Note that the proxy time series for VETF and LETF returns are \textit{only}
used when bootstrapping data sets providing the training/testing data
for the data-driven neural network approach followed in Section \ref{sec:Data-driven-Machine-Learning}.
In contrast, the parametric dynamics for the results of Section \ref{jump_section}
, Section \ref{sec:Jump-Diffusion results} and Section \ref{sec:Fixed weight simulations}
do not require historical proxy returns, since the LETF dynamics can
be obtained by calibrating equations (\ref{S_sde_jumps}) and (\ref{eq:dist})
to historical data as described in Appendix \ref{jump_params}.

\section{Parameters for jump diffusion model, fit to CRSP data 1926:1-2023:12.}
\label{jump_params}
The parameters for equations (\ref{S_sde_jumps}) and (\ref{eq:dist}) are fit
to the CRSP data, inflation adjusted,  with results in Table \ref{fit_params}.
We use the filtering technique described in \citep{mancini2009,contmancini2011,Dang2015a}
to estimate the parameters.

We also assume that the expense ratios are $c_\ell = .0089$ and $c_v = 0.0$,
and the LETF leverage factor is $\beta = 2$.

{\small
\begin{table}[hbt!]
\begin{center}
\begin{tabular}{ccccccc} \toprule[1pt]
 & $\mu$ & $\sigma$ & $\lambda$ & $p_{up}$ &
  $\eta_1$ & $\eta_2$  \\ \midrule
CRSP Index (real)       & 0.08732  & 0.1477&   0.3163  &  0.2258 &  4.3591& 5.5337 \\
\bottomrule[1pt]
\end{tabular}
\caption{Estimated annualized parameters for double exponential jump
diffusion model. Value-weighted CRSP index deflated
by the CPI. Sample period 1926:1 to 2023:12.
The average real return of a 30 day T-bill in the same period
was $r=0.0032$.
}
\label{fit_params}
\end{center}
\end{table}
}

\begin{singlespace}
\bibliographystyle{chicago}
\bibliography{paper}
\end{singlespace}

\end{document}